\documentclass[3p, times, twocolumn]{elsarticle}
\usepackage[utf8]{inputenc}
\usepackage[dvipsnames]{xcolor} 
\usepackage{natbib}
\usepackage{graphicx}
\usepackage[utf8]{inputenc}
\usepackage{multirow}
\usepackage{tabularx}
\usepackage{epsfig,color}
\biboptions{sort&compress}
\usepackage{amsmath,amssymb}
\usepackage{ulem}
\usepackage{soul}   
\usepackage{hyperref} 
\usepackage[version=4]{mhchem} 
\usepackage{verbatim}
\usepackage{lipsum} 
\usepackage{float}
\usepackage{geometry}
\usepackage{gensymb}
\usepackage{booktabs}
\usepackage{makecell} 
\usepackage{amssymb}

\begin{document}
\begin{frontmatter}
\title{Neutron reflectometry on superspreading and non-superspreading trisiloxane surfactants}

\author[1]{Joshua~Reed\corref{cor1}}
\ead{joshua.reed@pkm.tu-darmstadt.de}
\author[2]{Séforah Carolina Marques Silva}
\author[3]{Philipp Gutfreund}
\author[4]{Joachim~Venzmer\corref{cor1}}
\ead{joachim.venzmer@evonik.com}
\author[2]{Tatiana~Gambaryan‑Roisman}
\author[1]{Emanuel~Schneck\corref{cor1}}
\ead{emanuel.schneck@pkm.tu-darmstadt.de}
\address[1]{Institute for Condensed Matter Physics; TU Darmstadt; Hochschulstrasse 8; 64289 Darmstadt; Germany}
\address[2]{Institute for Technical Thermodynamics; TU Darmstadt; Peter-Grünberg-Str. 10; 64287 Darmstadt; Germany}
\address[3]{Institut Laue-Langevin; 71 Av. des Martyrs; 38000 Grenoble; France}
\address[4]{Research Interfacial Technology; Evonik Operations GmbH; Goldschmidtstr. 100; 45127 Essen; Germany}
\cortext[cor1]{Corresponding author}
\pagenumbering{arabic}
\noindent

\parindent=0cm
\setlength\arraycolsep{2pt}
\begin{abstract}\small 
Certain trisiloxane surfactants have the remarkable property of being able to superspread: Small volumes of water rapidly wet large areas of hydrophobic surfaces. The molecular properties of the surfactants which govern this technologically relevant effect are still under debate. To gain a deeper understanding, the surfactant behaviour during the spreading process needs to be studied at molecular length scales. Here, we present neutron reflectivity analyses of two trisiloxane surfactants of similar chemical structure, of which only one exhibits superspreading properties. We present an approach to determining the composition of the adsorbed surfactant layer in spread surfactant films at the solid-liquid interface, accounting for contributions from attenuated back-reflections of the neutron beam in films with thicknesses in the range of several tens to hundreds of micrometers. Differences between superspreading and non-superspreading surfactants with regard to their volume fraction profiles at the solid/liquid interface obtained in the self-consistent analysis of the reflectivity curves are in agreement with a simple explanation of the difference in spreading behaviour based on thermodynamics.
\end{abstract}
\end{frontmatter}

\newpage{}

\section{Introduction}
A number of trisiloxane surfactants exhibit the capability to cause small volumes of diluted aqueous solution to spread to surprisingly large diameters on sufficiently hydrophobic surfaces - a phenomenon known as superspreading. When placed on to a hydrophobic surface such as a polypropylene film, a small droplet of 50 \textmu l solution of 0.1 wt\% superspreader surfactant can be expected to spread to an area of 70-80 mm diameter at after 1 minute~\cite{Venzmer2011Aug}. This behaviour is not only limited to trisiloxane surfactants, but other surfactants also show this behaviour~\cite{kennedy1998organic}. Superspreading has also been observed in mixtures of cationic and anionic surfactants~\cite{Kovalchuk2015Dec}. This exceptional wetting ability has significant practical applications particularly as an agricultural adjuvant, where it ensures fast, uniform wetting of a plant's hydrophobic leaves, increased penetration of active ingredients into the plant, and allows for stomatal flooding all while lowering spray volumes by 30\%~\cite{BREAKTHRU.COM}.

Superspreading is well-documented, and is explained in terms of the system's free energy, expressed as the spreading coefficient, 
\begin{equation}
S = \gamma_{\text{S}} - (\gamma_{\text{SL}} + \gamma_{\text{L}}),
\label{eq:Youngs}
\end{equation}
where $\gamma_{\text{S}}$ is the surface tension of the bare solid, $\gamma_{\text{L}}$ is the tension of the surfactant solution's liquid surface, and $\gamma_{\text{SL}}$ is the interfacial tension between the solid and the solution. Spreading, i.e., complete wetting, occurs as long as $S$ is positive~\cite{Ross1992Mar}.

\begin{figure}[h]
	\centering
    \includegraphics[width=0.5\textwidth]{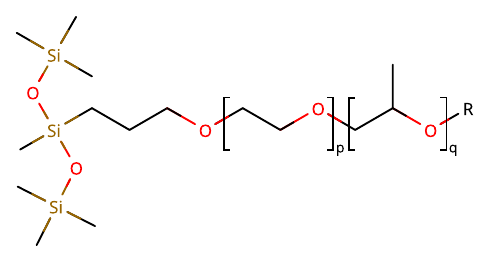}
	\caption{Chemical structures of S233 (p = 10, q = 2) and S240 (p = 6, q = 3). The structure consists of a hydrophobic trisiloxane group with a hydrophilic polyether chain whose monomer composition differs between the two molecules in terms of number of ethylene oxide (p) and propylene oxide units (q).}
	\label{fig:structure}
\end{figure} 

One such trisiloxane surfactant is BREAK-THRU\textsuperscript{\textregistered} S 240 (see Fig.~\ref{fig:structure}, abbreviated as S240 in the following), which has the structure M(D'R)M where M represents a trimethylsiloxy group, \ce{(CH3)3SiO1/2 -}, D'R represents \ce{-O1/2Si(CH3)(R)O1/2-}, and R represents a polyether chain consisting of a statistical mixture of ethylene oxide (p) and propylene oxide (q) units, in this case p = 6 and q = 3. Interestingly, there is a structurally quite similar trisiloxane surfactant, BREAK-THRU\textsuperscript{\textregistered} S 233 (abbreviated as S233, see also Fig.~\ref{fig:structure}), which does not exhibit superspreading. S233's polyether is longer and consists of more alkylene oxide units (p = 10, q = 2). Therefore, the difference in behaviour is dictated only by the R polyether monomer composition and length (see Fig.~\ref{fig:structure}). 

While considerable effort has been made to determine the exact mechanism of superspreading, the process remains challenging to elucidate due to the rapid speed of spreading, which makes real-time observation with molecular resolution difficult. Reflection anisotropy spectroscopy has provided some insights by comparing surfactant adsorption on hydrophobic substrates where it has revealed that superspreading S240 solutions exhibit anisotropic structures, while non-superspreading S233 behaves isotropically ~\cite{Williams2016Jan}. However, this technique was unable to determine the specific anisotropic structural differences.

Many studies have focused on superspreading trisiloxane surfactants such as S240 only, often attributing their superspreading properties to their compact T-shaped trisiloxane group. However, this will lead to oversimplified conclusions when ignoring the structurally similar non-superspreading trisiloxane surfactants. Efforts to compare these surfactants directly have yielded valuable insights. Wang et al., for example, used high-speed imaging to investigate the wetting kinetics of trisiloxane surfactants and other organic surfactants, such as C$_{12}$TAB and SDS~\cite{Wang2013Dec, Wang2015Dec, Wang2016Aug}. Kovalchuk et al. slowed down the spreading process by adding glycerol to the surfactant solution, allowing for more detailed observations~\cite{Kovalchuk2019May}.

Simulations and theoretical models have shed further light on the mechanics of superspreading. Computational studies, while resource-intensive due to the need to capture molecular-level interactions over macroscale timeframes, suggest that superspreading involves a unique interfacial behaviour. Large-scale molecular dynamics simulations by Isele-Holder et al. indicate that the three-phase contact line (TPCL) in superspreading systems exhibits a "rolling" transition rather than a sharp corner. This rolling transition minimises energy at the interface and supports the rapid movement of the TPCL at approximately 500~\textmu m/s~\cite{Isele-Holder2015May}. This behaviour contrasts with non-superspreading systems, where a fixed contact angle limits motion.

Marangoni flow has often been proposed as the driving mechanism for superspreading, with many authors attributing it to surface tension gradients along the droplet surface~\cite{Nikolov2002Feb, Chengara2007May, Nikolov2015Aug, Kovalchuk2019May}. One commonly cited piece of evidence for this theory is the observation that superspreading is most effective at a surfactant concentration of 0.1\%, while higher or lower concentrations are less effective, suggesting a role for Marangoni-driven redistribution. However, it is well-known for decades that this peculiar concentration dependence ("a lot does not help a lot") is not present when preventing evaporation. Also, there is sufficient experimental evidence that surface tension gradients do not play a role in superspreading~\cite{Wang2016Aug}. While Marangoni effects might contribute to the initial redistribution of surfactant molecules, they are unlikely to sustain the rapid and prolonged spreading characteristic of superspreading surfactants ~\cite{Venzmer2021Feb}.

An alternative hypothesis that we consider more convincing is that of a rolling mechanism, in which superspreading surfactants, like S240, form bilayers \cite{He1993PhaseTrisiloxanes} with minimal spontaneous curvature. These bilayers could "unzip" at the surface of the spreading drop, effectively transferring surfactant molecules from the bulk solution to the air-liquid interface. This unzipping action provides a continuous supply of surfactants to maintain rapid spreading, as illustrated in Fig.~6B in~\cite{Venzmer2021Feb}. Supporting evidence comes from foam film studies, which demonstrate distinct differences between superspreading and non-superspreading surfactants in their ability to form stable bilayer structures under dynamic conditions ~\cite{Sett2014Mar}. 

Irrespective of the specific spreading dynamics, it is important to acknowledge the gaps in our understanding of the equilibrium situation which is described by Eq.~\ref{eq:Youngs} and ultimately determines whether or not superspreading occurs. 

Both of these surfactants have almost identical liquid surface tensions at the typical spreading concentration of 0.1\% ($\gamma_{\text{L}}$~=~21.5~mN/m vs. $\gamma_{\text{L}}$~=~22.5~mN/m, respectively~\cite{Venzmer2024Aug}). Since $\gamma_{\text{S}}$ in Eq.~\ref{eq:Youngs} is surfactant-independent, the difference determining the sign of $S$ (and thus the occurrence of superspreading) must therefore be in the value of $\gamma_{\text{SL}}$. While $\gamma_{\text{SL}}$ is not directly accessible experimentally, it can however be safely assumed that this interfacial tension is governed by the organisation and coverage of surfactants adsorbed at the solid-liquid interface. This can be rationalised in the following way: the more a surfactant layer can minimise unfavourable contact of water and the hydrophobic surface, the lower $\gamma_{\text{SL}}$ will be. Consequently a method which can provide information about the molecular organisation at the interface is necessary.

Specular neutron reflectivity (NR) has been widely used to study the structure and organisation of lipid and surfactant layers at solid/liquid interfaces~\cite{lu1998neutron, plant1999supported, howse2001adsorbed}. The ability to resolve the structural detail of molecular size films down to nanometer spatial resolution makes it an ideal technique in studying the interfacial behaviour of these surfactant molecules by determining the laterally averaged density profiles perpendicular to the surface. In addition, NR is non-destructive and has the advantage of using selective deuteration or heavy water to create contrast between chemical components. The technique has already been applied to trisiloxane surfactants to study the temperature-dependent adsorption structure of trisiloxane surfactants on a titanium oxide surface~\cite{Gapon2023AprTrisiloxaneSurfactantsTiO}.

In the present work, the behaviour of superspreading and non-superspreading trisiloxane surfactants at the solid/liquid interface is systematically characterised using NR with the aim of shedding light on the molecular organisation in such films. Measuring neutron reflectivity on any liquid after spreading is difficult due to the nature of a circular puddle not being able to fully cover the rectangular neutron beam footprint. Furthermore, the superspreading results in films that are too thin to neglect the back-reflection from the upper surface and at the same time too thick to neglect beam attenuation. Accounting for attenuated reflectivity contributions from the air-water interface on the top-side of the film is therefore required when analysing the reflectivity data. 

\section{Results}
This study systematically characterises the behaviour of trisiloxane surfactants at the solid-liquid interface, alongside complementary measurements of their behaviour at the air-water interface. Building on the hypothesis introduced earlier, we aim to clarify the role of surfactant interfacial organisation, particularly at the solid interface, in determining the spreading behaviour of superspreading and non-superspreading surfactants. To prepare the solid surfaces for superspreading, the silicon blocks were hydrophobically functionalised with chlorotrimethylsilane (CTMS) by creating a molecularly thin layer of organosilyl groups from -\ce{Si(CH3)3} attached to the surface, thereby increasing the water contact angle to $\approx$~77$^{\circ}$. All experiments involving solid substrates were performed under conditions of saturated humidity, achieved with a large water bath (\ce{D2O}) surrounding the sample inside a closed measurement chamber. An illustration of this set-up can be found in the Supporting Information (Fig.~S1).\\
The reflectivity curves obtained with superspread and non-superspread droplets were modeled together with reference reflectivity measurements of bare solid surfaces and surfactant-loaded air/water interfaces in a self-consistent manner involving a number of common parameters. To maximize comprehensibility, the results are presented starting from the simplest (reference) configurations and then moving to the configurations that are more complicated to describe.     

\subsection{Reference Measurements}
\begin{figure}[h]
	\centering
        	\includegraphics[width=0.5\textwidth]{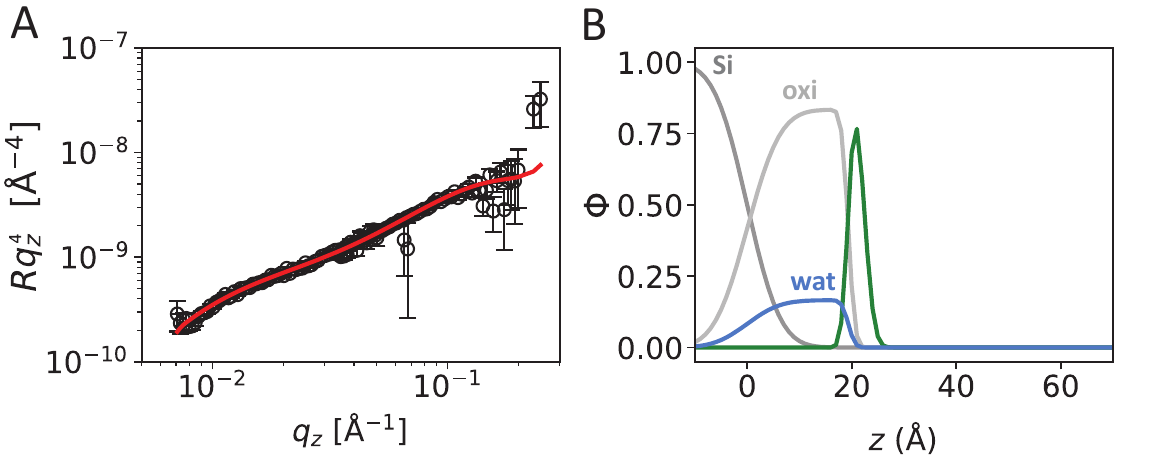}
    \includegraphics[width=0.5\textwidth]{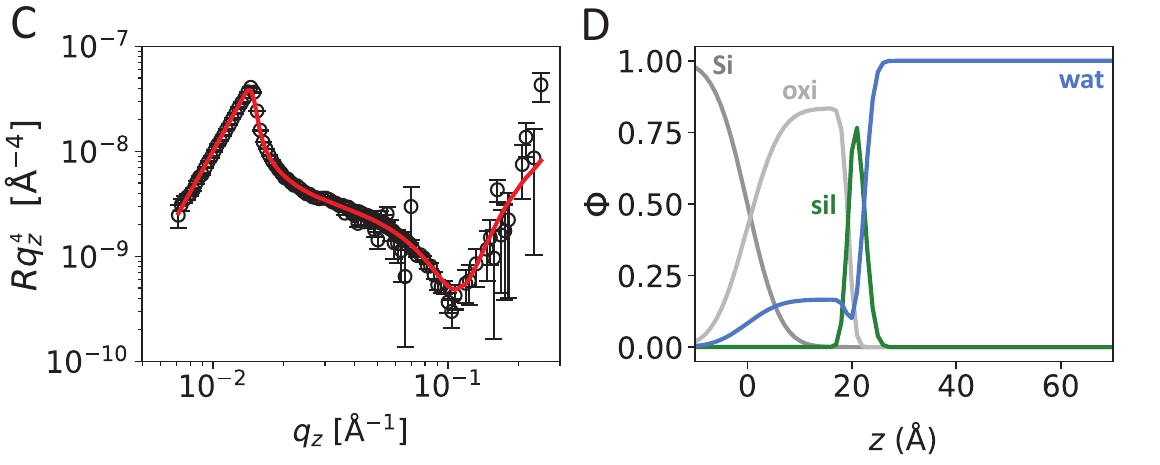}
    \includegraphics[width=0.5\textwidth]{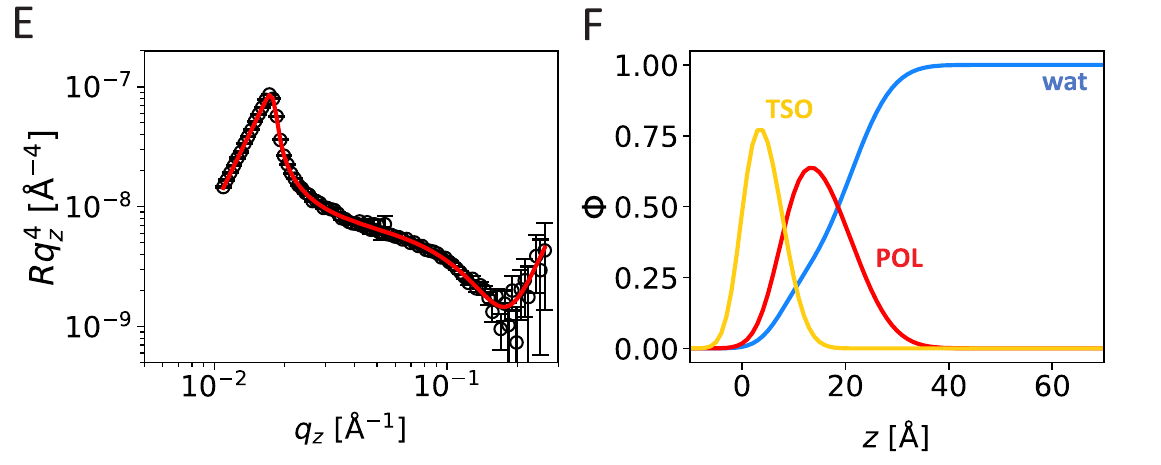}
	\caption{Neutron reflectivity data (A, C, E) and deduced volume fraction profiles (B, D, F) of the reference systems: (A, B) Bare silanized silicon block in air, (C, D) bare silanized silicon block in water, and (E, F) air-water interface of a 0.1\% surfactant solution (shown is S240). Solid lines in panels (A, C, E) indicate the best fits to the data that correspond to the volume fraction profiles in panels (B, D, F).}
	\label{fig:refl_references} 
\end{figure}

Newly silanized blocks were characterised by NR first, to determine a number of basic reference parameters for the models used in subsequent analyses. These reflectivity curves, obtained in air (close to 100\% humidity) and under water, are shown in Fig.~\ref{fig:refl_references}~A and C, respectively with solid red lines indicating the best-matching fits which provide the volume fraction profiles shown alongside. The reflectivity curves in air and under water were modelled with a common parameter set using a volume-fraction-based approach in the spirit of our earlier work~\cite{Schneck2015Apr, Rodriguez-Loureiro2017}. In this model, the SLD profile as a function of depth $z$, $\rho(z)$, is calculated from the volume fraction profiles of all chemical components $\phi_j(z)$ and from their SLDs $\rho_j$, where $j$ identifies the chemical component:  

\begin{equation}
\rho(z) = \rho_{\text{Si}} \phi_{\text{Si}}(z) + \rho_{\text{oxi}} \phi_{\text{oxi}}(z) + \rho_{\text{sil}} \phi_{\text{sil}}(z) + \rho_{\text{wat}}\phi_{\text{wat}}(z).
\label{sld:siwat}
\end{equation}

The last term, $\rho_{\text{wat}}\phi_{\text{wat}}(z)$, only applies to the measurement under water because air has negligible SLD. In this case, $\phi_{\text{wat}}(z)$ follows from the condition that the sum of all volume fractions amounts to 1 at each $z$-position:

\begin{equation}
\label{eq:spacefilling}
\sum_j\phi_j(z)\equiv 1.
\end{equation}

The silicon substrate ($j$ = 'Si') is modelled as a semi-infinite continuum with a constant SLD $\rho_{\text{Si}}$. The oxide layer ($j$ = 'oxi') was represented as a rough homogeneous layer with SLD $\rho_{\text{oxi}}$ and was allowed to have a finite water fraction $\Phi^\text{hydr}_\text{oxi}$ which resulted in a value of 17\% after fitting, which is the case for silicon blocks under water and in water-saturated air. The thickness $d_\text{oxi} \approx 16-19$~\AA~and the Si/oxide roughness obtained in the fit was $\sigma_\text{Si/oxi} \approx 5~ \AA$, in agreement with earlier work ~\cite{Rodriguez-Loureiro2017, micciulla2018versatile}. Silane ($j$ = 'sil') was also represented as a rough homogeneous slab. Its thicknesses and roughnesses were obtained in the fit as $d_\text{sil} \approx 3.5~\AA$, $\sigma_\text{oxi/sil} \approx 1.5 \AA$, and $\sigma_\text{sil/air}=\sigma_\text{sil/wat}\approx 1.5 \AA$. These values agree with the expected molecularly thin organosilyl layer. The volume fraction profiles corresponding to the solid surfaces in air and under water are shown in Fig.~\ref{fig:refl_references}~B and D, respectively.\\  
The third reference examined the air-water interface of a 0.1\% surfactant solution using a Langmuir trough. The corresponding SLD profile is:
\begin{equation}
\rho(z)=\rho_{\text{TSO}} \phi_{\text{TSO}}(z) + \rho_{\text{POL}} \phi_{\text{POL}}(z) + \rho_{\text{wat}} \phi_{\text{wat}}(z).
\label{sld:airwat}
\end{equation}
The adsorbed surfactants were modelled with two rough slabs in direct contact, a hydrophobic trisiloxane group ($j$ = 'TSO') and a hydrophilic polyether chain ($j$ = 'POL') as shown in Fig.~\ref{fig:structure}. These had adjustable thicknesses ($d_\text{TSO}$ and $d_\text{POL}$) and theoretical maximal volume fractions, $\Phi^{0}_{\text{TSO}}$ and $\Phi^{0}_{\text{POL}}$, valid in the limit of negligible interface roughness. The associated water fractions are being assumed to occupy any remaining space up to the TSO/POL interface as shown in the corresponding volume profiles in Fig.~\ref{fig:refl_references}~F, so that the water fraction in the polyether slab is $\Phi_{\text{wat}}^{\text{POL}}=1-\Phi_{0}^{\text{POL}}$. The maximum volume fractions of TSO and POL after accounting for interfacial roughness, denoted as $\Phi_{\text{max}}^{\text{TSO}}$ and $\Phi_{\text{max}}^{\text{POL}}$, are slightly lower than $\Phi^{0}_{\text{TSO}}$ and $\Phi^{0}_{\text{POL}}$ and can be visually extracted from the volume fraction profile in the figure. The parameters of $\phi_{\text{TSO}}(z)$ and $\phi_{\text{POL}}(z)$ were not entirely independent but constrained such that each of the two moieties occupies an overall volume consistent with the ratio of their known molecular volumes, which are given in Table~\ref{tab:volsld}. The scattering length densities $\rho_{\text{TSO}}$ and $\rho_{\text{POL}}$ were fixed at the calculated values, also shown in Table~\ref{tab:volsld}. 

For the superspreading S240, the fits yield thicknesses of $d_\text{TSO}\approx$ 7.5~\AA~and $d_\text{POL}\approx$ 13.5~\AA. Both values align with expectations considering, on one side, the diameter of the \ce{Si(CH3)3} groups and, on the other side, a polyether with 9 repeat units (p = 6 and q = 3) of $\approx$~4~\AA~length assuming a considerably contracted, quasi-random conformation. In the volume fraction profiles in Fig.~\ref{fig:refl_references}~F the surfactants are shown to form a compact layer with the hydrophobic moiety reaching $\Phi_{\text{max}}^{\text{TSO}}\approx$~74\% and the hydrophilic chain reaching $\Phi_{\text{max}}^{\text{POL}}\approx$~63\%. The two interfaces involving TSO are relatively sharp, but the transition of the polyether layer towards water is rather gradual, which reflects that this is not a dense layer but a bit like a hydrated polymer brush, for which similar profiles have been reported, albeit typically at even higher hydration levels~\cite{Schneck2015Apr, Rodriguez-Loureiro2017, micciulla2018versatile, Micciulla2023polymer}. The key parameters of the surfactant layers at the air/water interface are summarized in Table~\ref{tab:surfactant_params}. For the non-superspreader S233 shown in the Supporting Information (Fig.~S2) we obtain the same TSO thickness ($d_\text{TSO}^\text{S233} \approx$ 7.5~\AA) and comparable values of $\Phi_{\text{max}}^{\text{TSO}}$ and $\Phi_{\text{max}}^{\text{POL}}$, but the polyether profile is significantly more extended ($d_\text{POL}^\text{S233} \approx$ 16.0~\AA), as expected for a slightly longer polyether contour length (12 units, p = 10 and q = 2).

The surfactant-loaded liquid/air interfaces were also investigated with total-reflection X-ray fluorescence (TRXF) and grazing-incidence X-ray off-specular scattering (GIXOS)~\cite{Seeck:vv5026, Shen_2022, mukhina2022ph, mortara2024anion, mora2004x, o2005observation, PusterlaNano2022, Brezesinski2019Jul, Grava2023Oct}. The results and a more detailed discussion can be found in the Supporting Information. Most notably, TRXF revealed a slightly higher packing density (~8\% higher) for S240 than for S233.

\begin{table}[!ht]
    \renewcommand{\arraystretch}{1.5} 
    \centering
    \begin{tabular}{lcc}
    \toprule
    & \multicolumn{2}{c}{\textbf{Surfactant}} \\
    \textbf{Parameter} & \textbf{S240} & \textbf{S233} \\
    \midrule\midrule
    $d_\text{TSO}$ ($\pm 1.0$) & 7.5~\AA & 7.5~\AA \\ 
    $d_\text{POL}$ ($\pm 1.0$) & 13.5~\AA & 16.0~\AA \\ 
    \midrule
    \multicolumn{3}{c}{\textbf{Air/Water Interface}} \\ 
    \midrule
    $\Phi_{\text{POL}}^{\text{0}}$ ($\pm 0.05$) & 0.77 & 0.81 \\  
    $\Phi_{\text{POL}}^{\text{max}}$ ($\pm 0.05$) & 0.63 & 0.70 \\  
    $\Phi_{\text{POL}}^{\text{wat}}$ ($\pm 0.05$) & 0.23 & 0.19 \\  
    \midrule
    \multicolumn{3}{c}{\textbf{Solid/Liquid Interface}} \\ 
    \midrule
    $\Phi_{\text{POL}}^{\text{0}}$ ($\pm 0.05$) & 0.77 & 0.67\\ 
    $\Phi_{\text{POL}}^{\text{max}}$ ($\pm 0.05$) & 0.63 & 0.59 \\ 
    $\Phi_{\text{POL}}^{\text{wat}}$ ($\pm 0.05$) & 0.23 & 0.34 \\ 
    $\Phi_{\text{TSO}}^{\text{0}}$ ($\pm 0.05$) & 1.00 & 0.83 \\ 
    $\Phi_{\text{TSO}}^{\text{max}}$ ($\pm 0.05$) & 0.83 & 0.67 \\ 
    $\Phi_{\text{TSO}}^{\text{wat}}$ ($\pm 0.05$) & 0.00 & 0.17 \\ 
    \bottomrule
    \end{tabular}
    \caption{Parameters for the surfactants S240 and S233 obtained from fits of the reflectivity data at the air/water and solid/liquid interfaces. $\Phi_{j}^{\text{0}}$ is the maximal volume fraction of layer $j$ under hypothetical "no-roughness" conditions, $\Phi_{j}^{\text{max}}$ is its maximal volume fraction after roughness is applied, $\Phi_{j}^{\text{wat}}$ is its water fraction, and $d_j$ is its thickness parameter.}
    \renewcommand{\arraystretch}{1} 
    \label{tab:surfactant_params}
\end{table}

\subsection{S240 layers adsorbed to the solid/solution interface}
\begin{figure*}[t]
	\centering
	\includegraphics[width=0.8\textwidth]{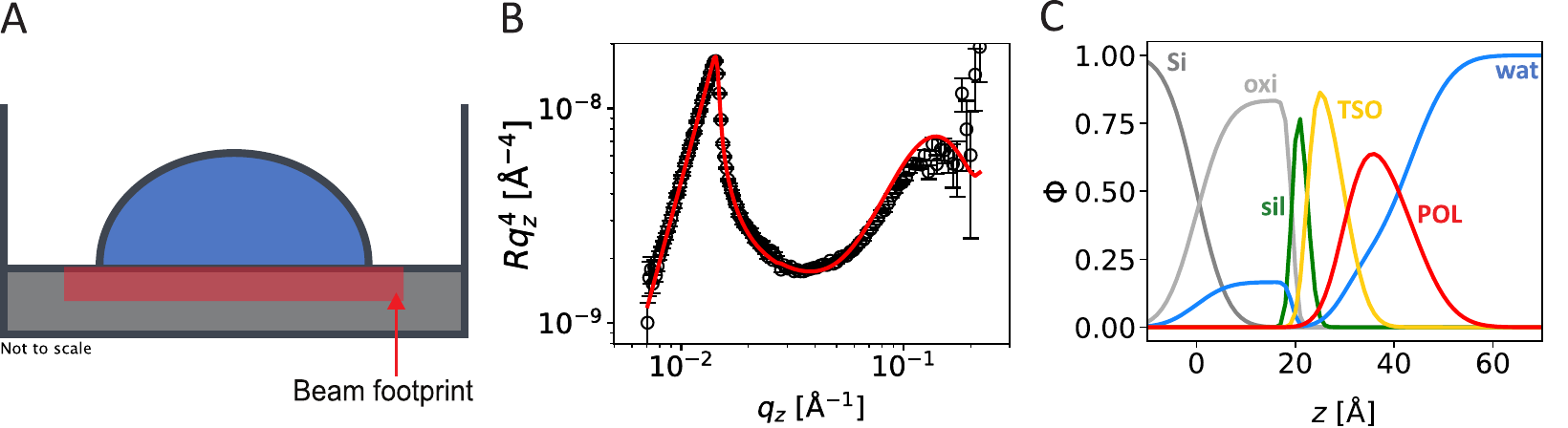}
	\caption{Demonstration of a "thick" droplet of S240 solution obtained by the addition of 200~\textmu l \ce{D2O} to a superspread puddle of S240 (0.1\% 10~\textmu l) (A). Neutron reflectivity data (B) and deduced volume fraction profiles (C) of this system. Solid lines in panels (B) indicate the best fits to the data that correspond to the volume fraction profiles in panels (C).}
	\label{fig:refl_S240_thick} 
\end{figure*}

To isolate the structure of surfactant layers adsorbed to the the solid/liquid interface, we first measured reflectivity from a thick droplet of S240 solution obtained by addition of 200 ~\textmu l \ce{D2O} to a superspread puddle of S240 (0.1\% 10~\textmu l). The addition of \ce{D2O} did not lead to further spreading  because the total surfactant amount in the solution was insufficient. This scenario is illustrated in Fig. ~\ref{fig:refl_S240_thick}~A and simplifies the analysis because, in contrast to a thin film, one does not have to account for reflectivity contributions from the air/water interface, as discussed further below.\\
Fig.~\ref{fig:refl_S240_thick}~B shows the reflectivity curve obtained with this sample. Again, the solid red line indicates the simulated reflectivity curve corresponding to a volume-fraction-based model.\\
In this experiment, the droplet is smaller than the beam footprint, necessitating an adjustable surface coverage parameter, $x_\text{c}$, in the reflectivity model. The reflectivity is thus modelled as a superposition of covered and uncovered regions:
\begin{equation}
R(q_z)=x_{\text{c}}R_{\text{c}}(q_z)+\left(1-x_{\text{c}}\right)R_{\text{nc}}(q_z),
\label{eq:surface coverage}
\end{equation}
where $R_{\text{nc}}(q_z)$ represents the reflectivity of the non-covered silanised silicon block in air as shown in Fig.~\ref{fig:refl_S240_thick}~A and described with Eq.~\ref{sld:siwat} without the term $\rho_{\text{wat}}\phi_{\text{wat}}(z)$.\\ 
$R_{\text{c}}(q_z)$ represents the reflectivity of the covered regions. Here, we integrate Eqs.~\ref{sld:siwat} and~\ref{sld:airwat} into one expression for the description of the surfactant-loaded solid/solution interface: 

\begin{equation}
\begin{split}
\rho(z) & = \rho_{\text{Si}} \phi_{\text{Si}}(z) + \rho_{\text{oxi}} \phi_{\text{oxi}}(z) + \rho_{\text{sil}} \phi_{\text{sil}}(z)\\ 
& + \rho_{\text{TSO}} \phi_{\text{TSO}}(z) + \rho_{\text{POL}} \phi_{\text{POL}}(z) + \rho_{\text{wat}} \phi_{\text{wat}}(z).
\end{split}
\label{sld:sfc_ad}
\end{equation}

The volume fraction profiles of each component were described in the same way as described in the previous paragraphs, however several parameters associated with the surfactants were allowed to assume different values at the solid/liquid interface than at the air/water interface. These were the roughness parameters as well as the coupled parameters for the maximal volume fractions of TSO and POL, in order to generally allow for different surfactant coverages at the to different interface types. The water volume fraction profile was again obtained through the constraint of Eq.~\ref{eq:spacefilling}.

The model reproduces the experimental data well, as seen in Fig.~\ref{fig:refl_S240_thick}~B. The volume fraction profiles obtained from the fit are shown in Fig.~\ref{fig:refl_S240_thick}~C and the key parameters are summarized again in Table~\ref{tab:surfactant_params}. For S240 we observe a very densely packed surfactant layer. While the polyether layer is strongly hydrated, the trisiloxane forms a compact and water-free layer on the silanized surface, indicating a space-filling configuration of the hydrophobic moieties at the interface. Attempts to model the data with water present in the trisiloxane layer were less effective as detailed in the Supporting Information (Fig.~S4). 

\subsection{Superspread S240 film}
\begin{figure}[h]
	\centering
    \includegraphics[width=0.5\textwidth]{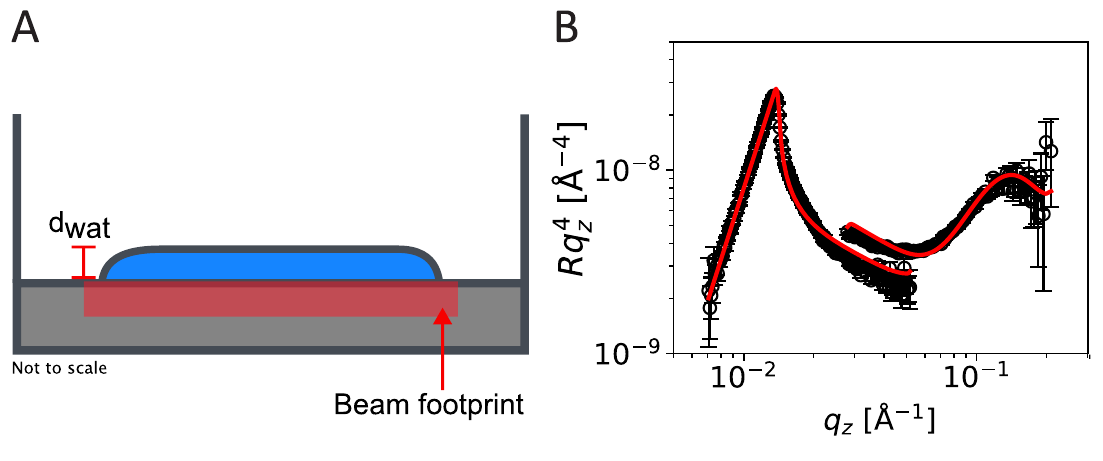}
	\caption{Demonstration of a "thin" droplet of a (0.1\% 10~\textmu l) superspread S240 solution(A). Neutron reflectivity data (B) of this system.}
	\label{fig:refl_S240_thin}
\end{figure}

A concentration of 0.1\% S240 solution is sufficient to cause superspreading of a deposited droplet. In this section we describe the reflectivity results of a 10 \textmu L 0.1\% solution that has superspread into a thin film on the hydrophobic substrate, as illustrated in Fig.~\ref{fig:refl_S240_thin}~A. The resulting film has a radius of approximately 15~mm, corresponding to a calculated thickness of 15~\textmu m using $d_{\text{W}}=V/(r^2\pi)$.

This thickness range of several tens of \textmu m presents experimental challenges, as the neutron reflectivity not only includes additional contributions from the liquid/air interface on the film's back side, but this contribution is also subject to attenuation effects, as noted earlier~\cite{browning2014specular}. As the film-internal path lengths, and thus the attenuation strengths, are different for the two incident angles $\theta_{\text{i}}$ used for the measurements, we are confronted with two non-overlapping reflectivity portions obtained at the two angles, as seen in Fig.~\ref{fig:refl_S240_thin}~B and Fig.~\ref{fig:refl_S233}~B, which have to be treated separately.\\ 
Taken together, modelling the reflectivity requires accounting for both covered and uncovered regions, as well as contributions from the liquid/air interface, whose attenuation depends on the incident angle $\theta_{\text{i}}$:

\begin{equation}
R(q_z, \theta_\text{i})=x_{\text{c}}R_{\text{c}}(q_z, \theta_\text{i})+\left(1-x_{\text{c}}\right)\cdot R_{\text{nc}}(q_z),
\label{eq:master}
\end{equation}
where
\begin{equation}
    R_{\text{c}}(q_z, \theta_\text{i})=R_{\text{SL}}(q_z)+T_{\text{SL}}(q_z)\cdot R_{\text{LA}}(q_z')\cdot T_{\text{WS}}(q_z')\cdot\alpha(q_z, \theta_\text{i}).
    \label{eq:reflcovered}
\end{equation}

$R_{\text{SL}}$ is the reflectivity of the solid/water interface (corresponding to the volume fraction profile in Fig.~\ref{fig:refl_S240_thick}~C), $T_{\text{SL}}$ is the transmittivity of the solid/water interface, and $R_{\text{LA}}$ is the reflectivity of the water/air interface (when starting from the water), which is derived from previously described air/water measurements, Fig.~\ref{fig:refl_references}~E and F. $T_{\text{WS}}$ is the transmittivity of the solid/water interface in the reverse direction. The attenuation effect is incorporated through a factor $\alpha(q_z, \theta_{\text{i}})$, as described in the Methods section. Finally, $q_z'$ denotes the refraction-corrected magnitude of the scattering vector that applies to the film-internal reflectivities and transmittivities,
\begin{equation}
q_z'=\sqrt{q_z^2-16\pi(\rho_{\text{wat}}-\rho_{\text{Si}})}.
\label{eq:qzprime}
\end{equation}

Since the back-side of a spread film is not necessarily as perfectly flat and smooth as a free air/water interface, $R_{\text{LA}}$ was not directly calculated from the SLD profiles deduced from the reference measurements at the free air water interface. Instead, we allowed for an elevated roughness through analytical convolution of the previously determined SLD profiles with a Gaussian function of adjustable width $\sigma_{\text{LA}}$.

The model of best fit reproduces the experimental reflectivity data relatively well, as shown in Fig.~\ref{fig:refl_S240_thin}~B. However, there are some discrepancies: the initial decay region does not match perfectly, and slight deviations are observed in the overlapping $q$ region. Nevertheless, the fit supports the methodology used.

From the spread amount of aqueous solution and the puddle's diameter, the average film thickness can be estimated to be about 15~\textmu m. This value is however inconsistent with the observed level of attenuation of the back reflection when assuming that the \ce{H2O} content is $\approx$~5\%, as reported by the position of the critical angle of total reflection, which probes the immediate vicinity of the interface. Instead, a larger "effective film thickness" of 58~\textmu m is required to reproduce the reflectivity data. Possible reasons for this discrepancy include a significantly higher \ce{H2O} content further away from the solid surface, heterogeneities in the water layer thickness, heterogeneities in the alignment of the backside with the solid surface, which contribute differently for the different incident angles having different footprints, as well as uncertainties in the exact relative scaling between first and second angles. Indeed, as shown in the Supporting Information (Fig. S6), agreement with the experimental data can be recovered with the nominal water layer thickness of 15~\textmu m when allowing for wavelength-dependent footprints for the two incident angles.

The additional roughness of the air/water interface was obtained as $\sigma_\text{LA}$~=~15~\AA, while overall the surface coverage was determined to be $x_{\text{c}}$~=~0.79. Moreover, it is crucial to emphasise that a water-free trisiloxane layer, as shown in Fig.~\ref{fig:refl_S240_thin}~C, at the solid-liquid interface was again necessary to reproduce the experimental data accurately. Attempts to model the system with water present in the trisiloxane layer were consistently ineffective as detailed in the Supporting Information. 

\subsection{Non-Superspreading Surfactant S233}
\begin{figure*}[t] 
	\centering
	\includegraphics[width=0.8\textwidth]{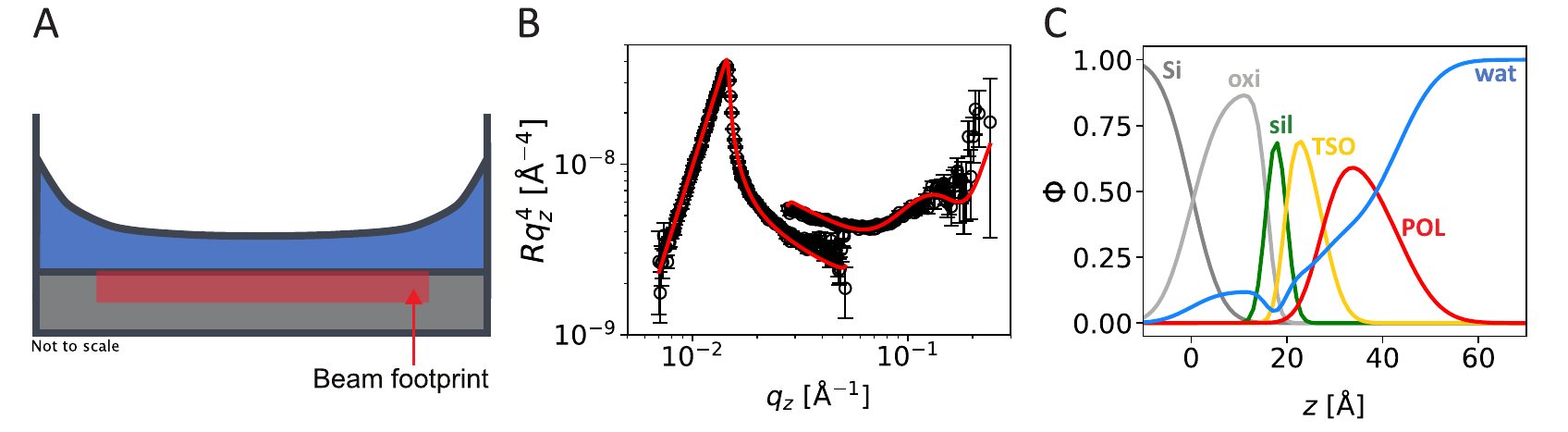}
	\caption{Demonstration of a "thick" droplet of non-superspread 0.1\% S233 solution (A). Neutron reflectivity data (B) and deduced volume fraction profiles (C) of this system. Solid lines in panels (B) indicate the best fits to the data that correspond to the volume fraction profiles in panels (C).}
	\label{fig:refl_S233}
\end{figure*}

Fig.~\ref{fig:refl_S233}~B and C show the reflectivity and volume fraction profiles for the non-superspreading surfactant S233. A larger volume of solution (0.1\% 500 \textmu l) was required to produce a sufficient amount of wetted area due to the significantly reduced spreading behaviour of S233. Therefore in this experiment, the entire silicon block was covered by S233 solution with the intention of making a simple system to measure only the solid-liquid interface.

However, as shown in Fig.~\ref{fig:refl_S233}~B, attenuated reflection contributions from the solution's back-side were clearly observed also in this experiment, as evidenced from the non-overlap of data points corresponding to the two different incident angles. The most likely reason for this result is that the wetting of the container walls resulted again in a rather thin water layer, as schematically depicted in Fig.~\ref{fig:refl_S233}~A, leading to the formation of a meniscus. Consequently, the reflectivity data for S233 were processed using the same methodology (Eqs.~\ref{eq:master}-\ref{eq:qzprime}) that was used for the S240 reflectivity data in Fig.~\ref{fig:refl_S240_thin}~B, involving a reference measurement from the air-water interface. Note, however, that $x_{\text{c}}$ was fixed at 1 for S233, because of the complete liquid coverage, such that the modelled reflectivity curve in Fig.~\ref{fig:refl_S233}~B is already fully described by Eq.~\ref{eq:reflcovered}. The model also fits well to the reflectivity data, as shown in Fig.~\ref{fig:refl_S233}~B. The obtained additional roughness of the water-air interface for S233, $\sigma_{\text{LA}}\approx$~6~\AA, is notably smaller than that for S240 ($\sigma_{\text{LA}}\approx$~15~\AA) in the thin film, which suggests that the characteristics of the surfactant-loaded backside interface of the thicker film are more similar to those of a meniscus-free macroscopic air/solution interface. The critical angle of total reflection in this measurement indicates a negligible \ce{H2O} fraction. With this assumption, the observed attenuation corresponds to a water layer thickness of $d_{\text{W}} \approx 230 \text{~\textmu m}$, which appears plausible.
The volume fraction profiles of the chemical components at the solid/liquid interface are shown in Fig.~\ref{fig:refl_S233}~C. The fits here follow the same constraints as in the previously discussed models. The water volume fraction profile was again obtained through the constraint of Eq.~\ref{eq:spacefilling}, and the surfactant thicknesses $d_\text{TSO}$ and $d_\text{POL}$ were again constrained to the same values for the air/water and solid/water interfaces. The obtained volume fraction profiles shown in Fig.~\ref{fig:refl_S233}~C reveal a maximal polyether volume fraction of $\Phi_\text{POL}^{\text{max}} = 0.58$ ($\Phi_\text{POL}^{0} = 0.67$) and a hydrophobic trisiloxane layer with a maximal volume fraction of $\Phi_\text{TSO}^{\text{max}} = 0.70$ ($\Phi_\text{TSO}^{0} = 0.83$). These values are considerably lower than at the air/water interface (see Table~\ref{tab:surfactant_params} and the Supporting Information, Fig.~S1), suggesting that S233 struggles much more than S240 to form dense layers at the solid surface. As a consequence, we find $\approx$ 17\% hydration ($\Phi_\text{TSO}^{\text{wat}}=1-\Phi_\text{TSO}^{0} = 0.17$) in the hydrophobic trisiloxane layer, which is in contrast to the water-free trisiloxane layer for S240. These findings are consistent with a second data set for S233, which is included in the Supporting Information (Fig.~S7). Attempts to fit the reflectivity data while forcing a water-free trisiloxane layer failed to converge on a suitable model, further supporting the observed hydration. Detailed comparisons of these alternative fits are also available in the Supporting Information.

\section{Discussion}
\begin{figure}[h]
	\centering
    \includegraphics[width=0.45\textwidth]{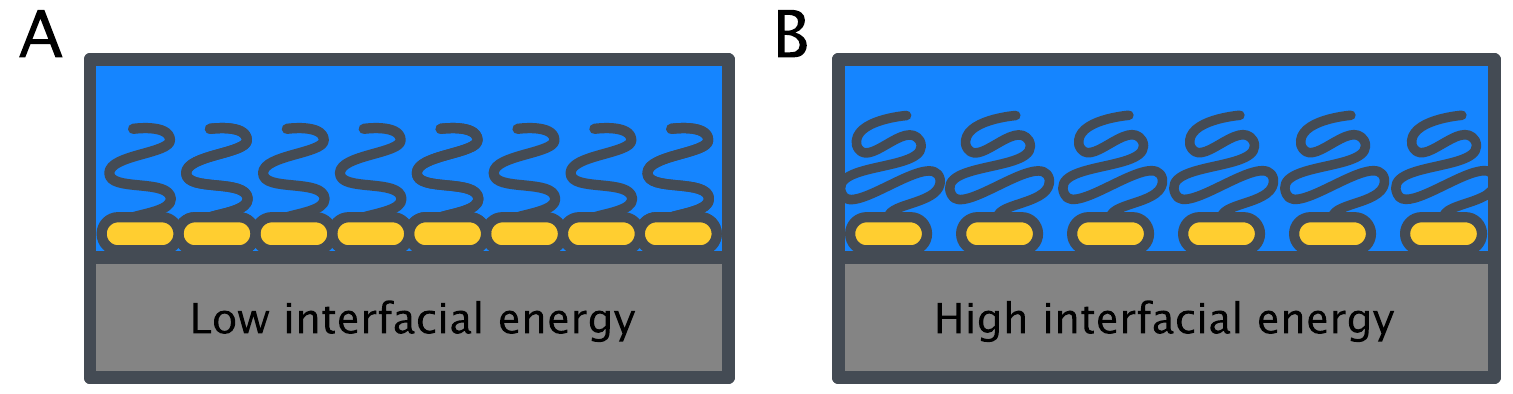}
  \caption{Visual representation of how adhered surfactant layers can affect volume of water in contact with the hydrophobic surface. A closely packed surfactant monolayer in (A) minimizes water contact with the hydrophobic surface, efficiently reducing the interfacial energy. Less dense packing or weak adsorption in (B) leads to a smaller reduction of interfacial energy.}
  \label{fig:packing}
\end{figure}

This study demonstrates that trisiloxane surfactants adsorbed at the hydrophobic solid-liquid interface form well-defined layers, comprising two distinct regions: an inner, poorly hydrated layer formed by the hydrophobic trisiloxane moiety and an outer, strongly hydrated layer formed by the hydrophilic polyether moiety. This organisation is summarised in Fig.~\ref{fig:packing}, which illustrates the structural differences between superspreading S240 and non-superspreading S233 surfactants. 

The clearest difference between S240 and S233 lies in the hydration of the hydrophobic trisiloxane layer directly in contact with the solid surface. For S240, the trisiloxane layer is water-free, indicating efficient packing that minimises water contact with the hydrophobic surface as seen in Fig.~\ref{fig:packing}~A. In contrast, S233 shows significant hydration within the trisiloxane layer, suggesting less efficient packing and the presence of water in unfavourable contact with the hydrophobic surface which can be seen in Fig.~\ref{fig:packing}~B. As also described in the Introduction and Eq.~\ref{eq:Youngs}, the more efficiently packed water-free trisiloxane layer of the S240 ensures a lower interfacial tension $\gamma_{\text{SL}}$ for the solid/liquid layer which leads to a positive spreading coefficient $S$, enabling superspreading. For S233, the hydrated TSO layer results in a higher $\gamma_{\text{SL}}$, flipping the sign of $S$ to negative and thereby only allowing partial wetting instead of superspreading. The inability of S233 to form a water-free trisiloxane layer is in line with a larger steric footprint of its polyether chain, which is also the reason behind the formation of spherical micelles in water \cite{He1993PhaseTrisiloxanes}. As shown in Fig.~\ref{fig:structure} and Table~\ref{tab:surfactant_params}, the polyether portion of S233 is slightly longer with 12 total monomer units (p = 10, q = 2) than that of S240's 9 units (p = 6 and q = 3), as reflected in the best fits. This increased length increases the volume requirement and also the area requirement of the polyether moiety, disrupting the efficient packing of the trisiloxane layer at the solid-liquid interface. Consequently, water cannot be prevented to penetrate into the interfacial region, coming into contact with the hydrophobic surface and increasing $\gamma_{\text{SL}}$.
On the other hand, S233 is almost equally good as S240 at efficiently packing at the air/water interface, as shown in the Supporting Information S2, which leads to the question of why it can do this at the air/water but not at the solid/water interface. One reason may be the higher tension of the bare air/water interface in comparison to that of the bare solid/water interface, so that dense molecular packing at the air/water interface may be free-energetically so favourable that it justifies entropically unfavourable stretching of the polyether chains. Another possible explanation may be based on differences between the two interface types with regard to the in-plane mobility of the adsorbed trisiloxane groups, but more research is needed to shed additional light on this question. For example, studying the behaviour of these surfactants at oil/water interfaces~\cite{kovalchuk2024spreading} can provide another interface to compare with, where we note that liquid/liquid interfaces are nowadays accessible also to NR~\cite{scoppola2016solvent}. Complementary insights may further be gained through interpretation of the adsorption kinetics and the surface tension isotherms of these surfactants \cite{Silva2023Apr}, which can reveal the occupation of different configurational states~\cite{fainerman2021multistate}.   Finally, molecular dynamics simulations could provide more detailed insights into the molecular structure and behaviour of the surfactant layers~\cite{kanduvc2024molecular}, while a combination of neutron reflectometry and X-ray reflectometry~\cite{schneck2024experimental} could help further refine the structural models developed in this work.

\section{Conclusions} 
This study characterised the interfacial behaviour of superspreading and non-superspreading trisiloxane surfactants using neutron reflectometry. The results highlight key differences between S240 and S233, particularly in how polyether chain variations affect hydration at the solid-liquid interface and determine superspreading ability. S240 forms a compact, water-free hydrophobic layer, while S233 cannot prevent partial hydration of the hydrophobic substrate due to less efficient packing and therefore less reduction of interfacial energy. These findings reinforce the link between molecular organisation, interfacial energy, and spreading behaviour. While challenges remain, such as handling attenuation effects in NR for thin layers, this work demonstrates the utility of NR in probing surfactant behaviour with molecular precision.

\section{Experimental section}
\subsection{Chemicals and Sample Preparation}
All chemicals were purchased from Merck (Darmstadt, Germany) and used as received without any further purification, unless otherwise stated. In addition, \ce{H2O} was ultra pure Milli-Q water (18 M$\Omega$cm$^{-1}$) in this study. BREAK-THRU\textsuperscript{\textregistered} S 233 and BREAK-THRU\textsuperscript{\textregistered} S 240 were provided by Evonik Operations GmbH (Essen, Germany), the chemical structures of these surfactants are shown in Fig.~\ref{fig:structure}. Surfactant solutions were prepared immediately before use by mixing 0.1 wt\% of surfactant and \ce{D2O} water with light shaking to the desired concentration (0.1\%). Silicon single crystal blocks (5~cm x 5~cm x 1~cm), purchased from Korth Kristalle (Altenholz, Germany), were polished on both sides and had a thin layer of native oxide (\ce{SiO2}) on their surface. These were functionalised to be hydrophobic prior to the experiments by firstly cleaning with a solvent cascade of chloroform (99.8\%), acetone (99.8\%), ethanol (99.9\%), and pure water, for 15 minutes each, then dried with nitrogen (\ce{N2}) gas and UV-ozone treated for 25 min. Secondly, the functionalisation was done by placing them in a sealed \ce{N2} environment with 15 ml beaker of chlorotrimethylsilane (CTMS, $\geq$ 98\%) and left overnight to vapour deposit, then washed with pure water when removed~\cite{Zhao2020Apr}. Surfactant solutions were deposited onto the block surface by placing the tip of an Eppendorf pipette, or for nano-quantities, the NanoLiter2020 (WPI Instruments, Friedberg, Germany) close to the surface and depositing gently the liquid in a single droplet.\\ 

\subsection{Neutron Reflectometry Experiments}
Specular neutron reflectometry (NR) was performed on the horizontal time-of-flight reflectometer FIGARO at the Institut Laue-Langevin (Grenoble, France)~\cite{campbell2011figaro}. In the experiments, the incident beam reaches the interface with an adjustable incident angle $\theta$. The reflectivity, i.e., the intensity ratio $R$ between reflected and incident beams, is recorded as a function of the scattering
vector component perpendicular to the interface, $q_z=(4\pi/\lambda)\sin{(\theta)}$, where $\lambda$ is the neutron wavelength. The reflectivity $R(q_z)$ is imposed by the scattering length density (SLD) profile $\rho(z)$ along the direction perpendicular to the interface, $z$. The SLD profile, in turn, originates from the interfacial distributions of all chemical components (see Results section) having their characteristic SLDs, where $b_k$ is the coherent scattering length of atomic nuclei of type $k$ and $N_k^i$ the number of such nuclei in the chemical component $i$ occupying the volume $v_i$.

\begin{equation}
\rho_i=\frac{1}{v_i}\sum_k N_k^ib_k.
\label{equation:sld}
\end{equation}

With that, the depth distribution of the molecular constituents in a sample can be reconstructed from the analysis of the $R(q_z)$ curves. In order to maximize the SLD contrast and thus the reflected intensity, deuterated water (D$_2$O, $\rho_{\text{D2O}}$~=~6.35$\times$10$^{-6}$~\AA$^{-2}$) was used in addition to H$_2$O ($\rho_{\text{H2O}}$~=~-0.56$\times$10$^{-6}$~\AA$^{-2}$).\\

\textbf{Solid surfaces:} All experiments involving solid surfaces were conducted in an air-tight sample cell with internally heated water reservoirs allowing for humidity control. Both humidity and temperature were monitored in real time using a sensor (B+B Thermo-Technik, Donaueschingen) with relative humidity kept above 95\% and temperature around room temperature. 
The measurements were carried out using two incident angles, $\theta_1$ = 0.71\degree~and $\theta_2$ = 2.41\degree~with a wavelength range of $2 \text{ \AA} < \lambda < 22 \text{ \AA}$ and the full width at half maximum $q_z$-resolution, $\Delta q_z/q_z$, was $q-z$-dependent and ranged between 4\% and 12\%.\\

\textbf{Liquid surfaces:} Measurements were performed at room temperature on bulk solutions of 0.1 w\% surfactant in both \ce{D2O} and air contrast matched water (ACMW, 92:8 \ce{D2O}/\ce{H2O} v/v, $\rho_\text{ACMW}$ = 0) inside a Langmuir trough installed at Figaro. The neutron beam approached through the air at two incident angles, $\theta_1$ = 0.72\degree and $\theta_2$ = 3.86\degree with a wavelength range of $2 \text{ \AA} < \lambda <  20 \text{ \AA}$ and  $\frac{\Delta q_z}{q_z} \approx 7-8\%$.\\

\textbf{Scattering length densities:} The SLD values of all chemical components (excluding water) are pre-established and constant. These are silicon ($\rho_{\text{Si}} = 2.07 \times 10^{-6} \text{\AA}^{-2}$), silicon oxide ($\rho_{\text{oxi}} = 3.47 \times 10^{-6} \text{\AA}^{-2}$) and silane ($\rho_{\text{sil}} = -0.30 \times 10^{-6} \text{\AA}^{-2}$, approximated from the known mass density of \ce{HSi(CH3)3} (0.635~g/cm$^3$)~\cite{trimethylsilane-GESTIS-Stoffdatenbank} and its coherent scattering length when accounting for the hydrogen not present in the covalently bound silane layer. Regarding the surfactants, we distinguish between their hydrophobic (trisiloxane, TSO) and hydrophilic (polyether, POL) parts for modelling purposes. Their SLDs were calculated as described in the Supporting Information, based on independent densiometry experiments using the allylpolyethers (ALP0620 for S233, ALP0540 for S240). The calculated SLDs are shown in Table~\ref{tab:volsld}.

\begin{table}[h]
\begin{center}
\caption{Summary of calculated molecular volumes ($V_{\text{i}}$) and scattering length densities ($\rho_{\text{i}}$) for the surfactant molecules and polyether chains.}
\label{tab:volsld}
\begin{tabular}{|l|c|c|}
\hline
Moiety & $V$ [nm$^3$] &  $\rho$ [10$^{-6}$~\AA$^{-2}$] \\
\hline
ALP0620 & 0.822 & 0.59 \\
ALP0540 & 0.664 & 0.47 \\
HMTS & 0.476 & -0.09 \\
\hline
S233 & 1.294 & 0.31 \\
S240 & 1.145 & 0.23 \\
\hline
\end{tabular}
\end{center}
\end{table}

\subsection{Calculation of theoretical reflectivity curves}
The SLD profiles $\rho(z)$ were discretised into thin layers of 2 \AA~thickness with constant SLD. The $q_z$-dependent reflection intensities were then determined by computing Fresnel’s reflection laws at each interface between the slabs and the using Parratt's iterative method for their phase-correct superposition~\cite{parratt1954surface}. 
To match the finite experimental $q_z$ resolution, the theoretical reflectivity curves were convoluted with Gaussian functions representing the experimental resolution. Finally, all unconstrained model parameters were adjusted to achieve the best agreement with all experimental reflectivity curves, characterised by the minimal chi-square deviation, $\chi^2$. Interface roughness parameters were constrained to have a lower limit of 1.5~\AA~and an upper limit of half the thickness of the slabs they belong to. The estimated parameter uncertainties are $\pm$~1~\AA~ for thicknesses and roughness parameters, $\pm$~0.05 for volume fractions parameters. These estimates are larger than the statistical uncertainty alone, because they also account for systematic uncertainties, which are typically the dominant contribution as discussed previously~\cite{Rodriguez-Loureiro2017}. 

\subsection{Attenuation of the neutron beam inside a thin water film}
For thin aqueous films of intermediate thicknesses in the range of several to several tens of micrometers, the reflection from the back side experiences significant attenuation that has to be taken into account when modelling the reflectivity curves. The attenuation factor $\alpha$ that applies to the back side reflection is given by the law of Lambert-Beer,
\begin{equation}
\alpha(q_z, \theta_\text{i}) = \mathrm{e}^{-L(q_z, \theta_\text{i})/L_\text{att}},
\label{eq:atten factor}
\end{equation}
and depends on the attenuation length $L_\text{att}$ and on the path length 
\begin{equation}
L(q_z, \theta_\text{i}) = 2d_\text{W}/\sin{\left[\theta(q_z, \theta_\text{i})\right]}, 
\label{eq:path length}
\end{equation}
where $\theta$ follows from $q_z$ and $\theta_\text{i}$ according to Snell's law, as shown in the Supporting Information.

\section{Supporting Material}
Experimental set-up; Reflectivity of S233 at the air/water interface; Synchrotron-based X-ray scattering and X-ray fluorescence measurements; Comparing different scenarios of water fractions in TSO for S240 and S233; Calculation of water-layer-internal incident angles; Accounting for an angle- and wavelength-dependent effectively probed surface coverage; Further neutron reflectivity measurements of a 0.1\% 200 $\mu$l sample of S233; Calculation of surfactant SLD using densiometry measurements; Tabulated neutron reflectivity fitting parameters.

\section{Author Contributions}
Joshua Reed: investigation, methodology, formal analysis, writing – original draft preparation; Séforah Carolina Marques Silva: investigation, formal analysis, writing – review and editing; Philipp Gutfreund: methodology, formal analysis, writing – review and editing; Joachim Venzmer: conceptualization, investigation, writing – original draft preparation, acquisition of funding, supervision; Tatiana Gambaryan-Roisman: conceptualization, investigation, writing – review and editing, acquisition of funding, supervision; Emanuel Schneck: conceptualization, methodology, investigation, formal analysis, writing – original draft preparation, acquisition of funding, supervision.

\section{Acknowledgements}
The authors would like to thank the Institut Laue-Langevin (ILL) for beam time allocation (DOI:10.5291/ILL-DATA.9-10-1740), and the ILL Soft Condensed Matter laboratories for their support. We would also like to thank Hacer Yalcinkaya for the densiometry measurements. The research leading to these results received funding from the European Union’s Horizon 2020 research and innovation program under the Marie Skłodowska-Curie grant agreement number 955612 (NanoPaInt).

\footnotesize
\setlength{\bibsep}{0pt}
\bibliographystyle{elsarticle-num}
\bibliography{bibliography}

\end{document}